\documentclass[twocolumn,twocolappendix]{aastex631}

\begin{document}

\title{Mass-gap Black Holes in Coalescing Neutron Star--Black Hole Binaries}

\author[0000-0002-0031-3029]{Zepei\,Xing}
\affiliation{Département d’Astronomie, Université de Genève, Chemin Pegasi 51, CH-1290 Versoix, Switzerland}
\affiliation{Gravitational Wave Science Center (GWSC), Université de Genève, CH1211 Geneva, Switzerland}
\affiliation{Center for Interdisciplinary Exploration and Research in Astrophysics (CIERA), 1800 Sherman, Evanston, IL 60201, USA}

\author[0000-0001-9236-5469]{Vicky Kalogera}
\affiliation{Department of Physics \& Astronomy, Northwestern University, 2145 Sheridan Road, Evanston, IL 60208, USA}
\affiliation{Center for Interdisciplinary Exploration and Research in Astrophysics (CIERA), 1800 Sherman, Evanston, IL 60201, USA}

\author[0000-0003-1474-1523]{Tassos\,Fragos}
\affiliation{Département d’Astronomie, Université de Genève, Chemin Pegasi 51, CH-1290 Versoix, Switzerland}
\affiliation{Gravitational Wave Science Center (GWSC), Université de Genève, CH1211 Geneva, Switzerland}

\author[0000-0001-5261-3923]{Jeff~J.~Andrews}
\affiliation{Department of Physics, University of Florida, 2001 Museum Rd, Gainesville, FL 32611, USA}

\author[0000-0002-3439-0321]{Simone\,S.\,Bavera}
\affiliation{Département d’Astronomie, Université de Genève, Chemin Pegasi 51, CH-1290 Versoix, Switzerland}
\affiliation{Gravitational Wave Science Center (GWSC), Université de Genève, CH1211 Geneva, Switzerland}

\author[0000-0002-6842-3021]{Max\,Briel}
\affiliation{Département d’Astronomie, Université de Genève, Chemin Pegasi 51, CH-1290 Versoix, Switzerland}
\affiliation{Gravitational Wave Science Center (GWSC), Université de Genève, CH1211 Geneva, Switzerland}

\author[0000-0001-6692-6410]{Seth\,Gossage}
\affiliation{Center for Interdisciplinary Exploration and Research in Astrophysics (CIERA), 1800 Sherman, Evanston, IL 60201, USA}

\author[0000-0003-3684-964X]{Konstantinos\,Kovlakas}
\affiliation{Institute of Space Sciences (ICE, CSIC), Campus UAB, Carrer de Magrans, 08193 Barcelona, Spain}
\affiliation{Institut d’Estudis Espacials de Catalunya (IEEC), Carrer Gran Capit\`a, 08034 Barcelona, Spain}

\author[0000-0001-9331-0400]{Matthias\,U.\,Kruckow}
\affiliation{Département d’Astronomie, Université de Genève, Chemin Pegasi 51, CH-1290 Versoix, Switzerland}
\affiliation{Gravitational Wave Science Center (GWSC), Université de Genève, CH1211 Geneva, Switzerland}

\author[0000-0003-4474-6528]{Kyle~Akira~Rocha}
\affiliation{Department of Physics \& Astronomy, Northwestern University, 2145 Sheridan Road, Evanston, IL 60208, USA}
\affiliation{Center for Interdisciplinary Exploration and Research in Astrophysics (CIERA), 1800 Sherman, Evanston, IL 60201, USA}

\author[0000-0001-9037-6180]{Meng\,Sun}
\affiliation{Center for Interdisciplinary Exploration and Research in Astrophysics (CIERA), 1800 Sherman, Evanston, IL 60201, USA}

\author[0000-0003-1749-6295]{Philipp\,M.\,Srivastava}
\affiliation{Center for Interdisciplinary Exploration and Research in Astrophysics (CIERA), 1800 Sherman, Evanston, IL 60201, USA}
\affiliation{Electrical and Computer Engineering, Northwestern University, 2145 Sheridan Road, Evanston, IL 60208, USA}

\author[0000-0002-7464-498X]{Emmanouil\,Zapartas}
\affiliation{Institute of Astrophysics, FORTH, N. Plastira 100,  Heraklion, 70013, Greece}



\begin{abstract}
The existence of a mass gap of $3-5\,M_{\odot}$ between the heaviest neutron stars (NSs) and the lightest black holes (BHs), inferred from the BH mass distribution in low-mass X-ray binaries (LMXBs), has been suggested for decades. The gravitational-wave (GW) source GW230529 is most likely a neutron star--black hole (NSBH) merger, with the BH mass falling within this gap. This detection strongly challenges the existence of the gap and has implications for the NSBH population, including a revised BH mass distribution and an updated local merger rate. In this study, we employ \texttt{POSYDON}, a binary population synthesis code that integrates detailed single- and binary-star models, to investigate coalescing NSBH binaries, focusing on the BH mass distribution of the intrinsic NSBH merger population.
For typical population models, we find that better matching the observed BH mass distribution requires the use of common-envelope efficiencies exceeding unity, a rather uncomfortable choice since most energy sources are already included. Alternatively, we find that a two-stage CE prescription calibrated to 1D hydrodynamic simulations \citep{2019ApJ...883L..45F} has a similar effect. Moreover, motivated by a possible explanation for the firm presence of the mass gap in the LMXB sample, we examine models where the NS birth mass is limited to be $\lesssim 2\,M_{\odot}$ and find excellent agreement with GW observations.
Additionally, we present observable NSBH merger property distributions, finding good agreement with the measured properties of observed systems and a predicted fraction of NSBH mergers with potential electromagnetic counterparts that ranges from $1\%$ all the way up to $32\%$.

\end{abstract}

\section{Introduction} \label{sec:intro}

The first observations consistent with masses for neutron star--black hole (NSBH) merger events were announced by \citet{2021ApJ...915L...5A} in 2021. One event, GW200105, 
has component masses $M_{\rm{BH}} = 8.9^{+1.2}_{-1.5}\,M_{\sun}$ and $M_{\rm{NS}} = 1.9^{+0.3}_{-0.2}\,M_{\sun}$, and an effective inspiral spin parameter $\chi_{\rm eff} = -0.01^{+0.11}_{-0.15}$, strongly peaking around zero. The other event, GW200115, has component masses $M_{\rm BH} = 5.7^{+1.8}_{-2.1}\,M_{\sun}$ and $M_{\rm NS} = 1.5^{+0.7}_{-0.3}\,M_{\sun}$, and effective inspiral spin parameter $\chi_{\rm eff} = -0.19^{+0.23}_{-0.35}$. The possible negative $\chi_{\rm eff}$ suggests that the BH spin is anti-aligned with the orbit (although \citet{2021ApJ...922L..14M} later argued that a nonspinning BH is more consistent with astrophysical understanding). From these two events, \citet{2021ApJ...915L...5A} inferred a local merger rate density of $\mathcal{R}_{\rm{NSBH}}= 45^{+75}_{-33}\,\rm{Gpc}^{-3}\rm{yr}^{-1}$. 

In April 2024, \citet{2024ApJ...970L..34A} reported the detection of GW230529, most likely a NSBH merger event, discovered during the fourth observing run of the LIGO, Virgo, and KAGRA (LVK) network. The primary mass of GW230519 is estimated to be $3.6^{+0.8}_{-1.2}\,M_{\odot}$ and the secondary mass is $1.4^{+0.6}_{-0.2}\,M_{\odot}$. The effective inspiral-spin $\chi_{\rm{eff}}$ is $-0.10^{+0.12}_{-0.17}$, which suggests either an anti-aligned spin component or negligible spins. The $\chi_{\rm{eff}}$ is correlated with the mass ratio $q = M_{\rm NS}/M_{\rm BH}$, with more negative $\chi_{\rm{eff}}$ corresponding to a more comparable mass ratio. Including the new event GW230529, \citet{2024ApJ...970L..34A} updated the inferred local NSBH merger rate density upwards to $\mathcal{R}_{\rm{NSBH}}= 94^{+109}_{-64}\,\rm{Gpc}^{-3}\rm{yr}^{-1}$.

Prior to gravitational wave (GW) observations, the BH mass distribution inferred from the X-ray binary populations indicated a low-mass boundary of $\simeq 5\,M_{\odot}$ \citep{1998ApJ...499..367B,2010ApJ...725.1918O,2011ApJ...741..103F}. In the meantime, the maximum NS mass has been suggested to be $\simeq 2.35\,M_{\odot}$ based on electromagnetic observations \citep{2022ApJ...934L..17R} and below $\simeq 2.59\,M_{\odot}$ from GW observations \citep{2020ApJ...892L...3A, 2020ApJ...896L..44A, 2024arXiv240711153C}. Theoretically, NS equation of state (EoS)  determines the mass-radius relation and sets a limit on the mass of nonrotating NSs, referred to as the Tolman–Oppenheimer–Volkoff limit $M_{\mathrm{TOV}}$ \citep{1939PhRv...55..364T,1939PhRv...55..374O}. $M_{\mathrm{TOV}}$ has been constrained to be $2-3\,M_{\odot}$ \citep{1974PhRvL..32..324R,1996ApJ...470L..61K,2001ApJ...550..426L,2018ApJ...852L..25R,2021ApJ...908..122G,2024PhRvD.109d3052F,2025PhRvD.111b3029G}. The lightest BH and the heaviest NS lead to a mass gap of $\sim 3-5\,M_{\odot}$. Recently, several candidate systems have been proposed to contain a BH with the mass overlapping the lower mass gap from electromagnetic observations \citep{2017ApJ...846..132H,2019MNRAS.488.1026Z,2022MNRAS.516.2023C,2022MNRAS.515.3105S,2023ApJ...944..165B,2024Sci...383..275B, 2024NatAs...8.1583W} and from GW detections \citep{2020ApJ...896L..44A,2021ApJ...923...14A,2023PhRvX..13d1039A,2024PhRvD.109b2001A}, but most have not been confirmed with high confidence. 

\citet{2012ApJ...749...91F} built two analytical supernova (SN) prescriptions based on different assumptions about the instability growth timescale to map the carbon-oxygen core mass to compact object remnant mass. With the rapid prescription, the mass gap can be generated, while the delayed prescription predicts a continuous remnant mass distribution. No theoretical constraints are able to determine which one is more representative. The detailed one-dimensional core-collapse models of \citet{2016ApJ...821...38S} allow for the formation of BHs around $4\,M_{\odot}$ but predicted very few lower-mass BHs due to the lack of fallback SNe that could produce them. 3D SN simulations have shown that fallback can lead to the formation of mass-gap BHs \citep{2018ApJ...852L..19C, 2020MNRAS.495.3751C, 2021ApJ...920L..17V}. Most recently, \citet{2024arXiv241207831B} presented results from 3D core-collapse simulations that link NS/BH masses, kicks, and spins to their progenitors and demonstrated the formation of BHs with masses in the gap. 

GW230529 is the first GW detection of a compact binary whose primary component is highly likely to fall within the mass gap, potentially resolving the longstanding debate on the existence of the mass gap. Assuming that the range of $3-5\,M_{\odot}$ represents the mass gap, \citet{2024ApJ...970L..34A} provided a merger rate for binaries with components within the gap of $\mathcal{R}_{\rm{gap}}= 24^{+28}_{-16}\,\rm{Gpc}^{-3}\rm{yr}^{-1}$ or $\mathcal{R}_{\rm{gap}}= 33^{+89}_{-29}\,\rm{Gpc}^{-3}\rm{yr}^{-1}$ based on two different population models.
This discovery along with those observed electromagnetically in detached binaries now provides strong evidence that BHs with masses in the claimed low-end mass gap actually do form in nature. Nevertheless the existence of a mass gap in the observed LMXB population is not under debate at present \citep{2010ApJ...725.1918O,2011ApJ...741..103F}, and consequently the question of what causes the LMXB mass gap becomes even more pressing \citep[see,][]{2001ApJ...554..548F,2012ApJ...757...36K}. Most recently, \citet{2023ApJ...954..212S} investigated selection biases against mass-gap BHs, taking into account that a dynamical BH mass measurement requires the existence of transient behaviors of the corresponding LMXBs. They found that, with rapid population synthesis simulations, all model combinations predict detection of LMXBs containing mass-gap BHs. However, using detailed binary stellar-evolution models, they found that observational biases against mass-gap BHs appear when the maximum NS birth mass is less than $\simeq 2\,M_{\odot}$. If the maximum NS birth mass is greater, the mass gap would be filled by BHs formed through accretion-induced collapse in LMXBs and we should have observed plenty of them. 

\citet{2024ApJ...974..211Z} investigated the formation of mass-gap NSBHs using the rapid binary population synthesis (BPS) code \texttt{COMPAS} \citep{2017NatCo...814906S,2022ApJS..258...34R}, examining local merger rates, compact object mass distributions, and the properties of potential electromagnetic counterparts (EMCs). In this study, we use the BPS code \texttt{POSYDON} \citep{2023ApJS..264...45F}, which incorporates extensive grids of detailed stellar binary-evolution models, to study the formation of coalescing NSBH systems across the Universe. Particularly, we focus on the BH mass distributions of coalescing  NSBHs and the effects of model variations of common envelope (CE) evolution and SN kicks. With \texttt{POSYDON}, we are also able to self-consistently estimate the BH spins and, thus, more accurately infer the fraction of associated EMCs.

\section{Methods} \label{sec:method}

To simulate populations of binaries, we employ an updated version (v2) of the publicly available code \texttt{POSYDON}\footnote{We utilized the version of the {\tt POSYDON} code identified by the commit hash 5acad13, available at \href{https://github.com/POSYDON-code/POSYDON/}{https://github.com/POSYDON-code/POSYDON/}, along with the {\tt POSYDON} v2 dataset that will be published here: \href{https://doi.org/10.5281/zenodo.15194708}{https://doi.org/10.5281/zenodo.15194708}.}\citep{2024arXiv241102376A}, which integrates detailed single- and binary-star model grids, simulated using the stellar evolution code Modules for Experiments in Stellar Astrophysics \citep[{\tt MESA},][]{2011ApJS..192....3P,2013ApJS..208....4P,2015ApJS..220...15P,2018ApJS..234...34P,2019ApJS..243...10P, 2023ApJS..265...15J}. Compared to \texttt{POSYDON} v1 \citep{2023ApJS..264...45F}, \texttt{POSYDON} v2 extends to include grids with eight metallicities $Z=10^{-4},10^{-3},10^{-2},0.1,0.2,0.45,1,2\,Z_{\odot}$. 

We evolve a population of $10^6$ binaries for each metallicity, starting from two zero-age main-sequence stars. We sample the primary mass following the initial mass function from \citet{2001MNRAS.322..231K}, within the range of $7-120\,M_{\odot}$. The secondary mass, ranging from $0.5\,M_{\odot}$ to $120\,M_{\odot}$, adheres to a flat distribution from the minimal ratio of $0.05$ to $1$ with respect to the primary mass. The initial orbital periods follow the distribution of \citet{2013A&A...550A.107S} from $1\,\rm{days}$ to $6000\,\rm{days}$. We extend this range to $0.35\,\rm{days}$ with a logarithmic uniform distribution. We assume the initial orbits are circular and the two stars are tidally synchronized. For all the population models presented in this study, we use the nearest-neighbor interpolation scheme to interpolate between models in the binary grids \citep[see Section~7 in][]{2023ApJS..264...45F}.

To calculate the compact object masses through core-collapse supernova (CCSN), we adopt the \citet{2012ApJ...749...91F} delayed prescription, which allows the formation of mass-gap BHs. Regarding the CCSN natal kick velocity for NSs, we adopt a Maxwellian distribution with a velocity dispersion of $\sigma_{\rm{CCSN}}=265\,\rm{km\,s^{-1}}$ \citep{2005MNRAS.360..974H}. Some studies have suggested that the NS kick should be weaker. For example, \citet{2023MNRAS.521.2504O} obtained a velocity dispersion of $\sigma_{\rm{CCSN}}=61.6\,\rm{km\,s^{-1}}$ from the populations of NS binaries. However, the authors carefully noted that this estimate should be considered as the lower limit because their sample excludes the disrupted systems in which the NSs receive large kicks. As a result, we also consider $\sigma_{\rm{CCSN}}=61.6\,\rm{km\,s^{-1}}$ and a moderate value of $\sigma_{\rm{CCSN}}=150\,\rm{km\,s^{-1}}$ for the purpose of parameter study. For electron-capture SN (ECSN), we adopt the prescription of \citet{2004ApJ...612.1044P}, and a velocity dispersion of $\sigma_{\rm{CCSN}}=20\,\rm{km\,s^{-1}}$ \citep{2019MNRAS.482.2234G}. BH kicks follow the same distribution but are rescaled by a factor of $1.4\,M_{\odot}/M_{\rm BH}$.

In \texttt{POSYDON}, the maximum NS mass is, by default, set to be $2.5\,M_{\odot}$. To account for selection biases against mass-gap BHs in LMXBs, we also consider a maximum NS birth mass of $M_{\rm{NS,birth-max}} = 2\,M_{\odot}$ \citep{2023ApJ...954..212S}. To achieve this, we assume that all compact objects born with masses above $2\,M_{\odot}$ as predicted by the delayed prescription are light BHs instead of NSs. This assumption excludes these binaries from the NSBH population. We acknowledge that we do not include a self-consistent SN prescription for classifying remnant type and predicting the masses of these specific compact objects. A small population of double NSs with one NS heavier than $2\,M_{\odot}$ at birth would become NSBHs based on this assumption about the NS maximum birth mass. In our simulated populations, the contribution of these systems is insignificant.

Utilizing the stellar profiles of the BH progenitors at carbon depletion, we calculate the BH spins by following the method described in Appendix D of \citet{2021A&A...647A.153B}. We assume that the BH spin vector is the same as that of its progenitor, and that the progenitor is always aligned with the orbit. We ignore any change in the BH spin direction due to mass transfer. The final orbital tilt is determined by the combination of two tilts originating from two SN kicks \citep{2024A&A...683A.144X}. We also take into account the change in BH spin magnitude due to accretion. However, it has a limited effect because we apply Eddington-limited accretion.  

For CE evolution, we employ the $\alpha-\lambda$ prescription \citep{1984ApJ...277..355W,1987A&A...183...47D,1988ApJ...329..764L}. The CE efficiency parameter $\alpha_{\rm{CE}}$ determines how efficient the energy can be utilized to expel the envelope. By default, we set $\alpha_{\rm{CE}} = 1$. The parameter $\lambda_{\rm{CE}}$ depends on the mass and the internal structure and energy profile of the stars at the onset of CE evolution \citep{2000A&A...360.1043D}. Unlike most rapid BPS codes, which use pre-tabulated $\lambda_{\rm{CE}}$ values obtained from studies on single stars, \texttt{POSYDON} allows for on-the-fly calculations of $\lambda_{\rm{CE}}$ based on the profiles of the stars. To determine the boundary between the helium core and the envelope, we select the layer where the hydrogen mass fraction $X_{\rm{H}} = 0.1$ by default \citep[see Section~8.2 in][]{2023ApJS..264...45F}. The binding energy of the envelope is highly sensitive to the location of the boundary because the deeper layers are the most tightly bound \citep{2000A&A...360.1043D,2013A&ARv..21...59I,2019ApJ...883L..45F}. As a result, we consider a variation of $X_{\rm{H}} = 0.3$ to define the core-envelope boundary in our population study. This choice follows the complete simulation of CE phase of \citet{2019ApJ...883L..45F} using MESA, which shows that the donor star begins to contract and the binary detaches when the donor star's hydrogen abundance drops to $\approx0.3$. \citet{2019ApJ...883L..45F} further illustrated that, after detachment, the donor star re-expands to initiate a highly non-conservative stable mass transfer phase. Accordingly, \texttt{POSYDON} treats CE evolution as a two-phase process. In the first phase, a rapid inspiral-in occurs and the layers outside the defined core-envelope boundary are subject to ejection. In the second phase, the binary undergoes fully non-conservative mass transfer until the donor star is stripped to its helium core and the transferred mass is lost from the vicinity of the accretor.

From the simulated binaries, we select those binaries forming a NS and a BH that merge within a Hubble time to comprise our synthetic population. Then, we distribute the synthetic population across the cosmic history of the Universe at every $\Delta t_{\rm{i}}$ ($100\,\rm{Myr}$ by default) cosmic time interval centered on the redshift $z_{\rm{i}}$ \citep[see Section 6.3 in,][]{2024arXiv241102376A}. To account for the metallicity- and redshift-dependent star formation history (SFH), for each coalescing NSBH system, we calculate its cosmological weight based on the star formation rate density and metallicity distribution from the TNG100 Illustris large-scale cosmological simulation \citep{2019ComAC...6....2N}. Then, the weighted population represents the intrinsic population. Finally, we calculate the detection probability $p_{\rm{det}}$ for each binary assuming a network of LIGO-Hanford, LIGO-Livingston, and Virgo at \texttt{design} sensitivity \citep[see][]{2020LRR....23....3A} to generate the observable populations.

\section{Results} \label{sec:result}

\begin{figure*}[t]
\includegraphics[width=\textwidth]{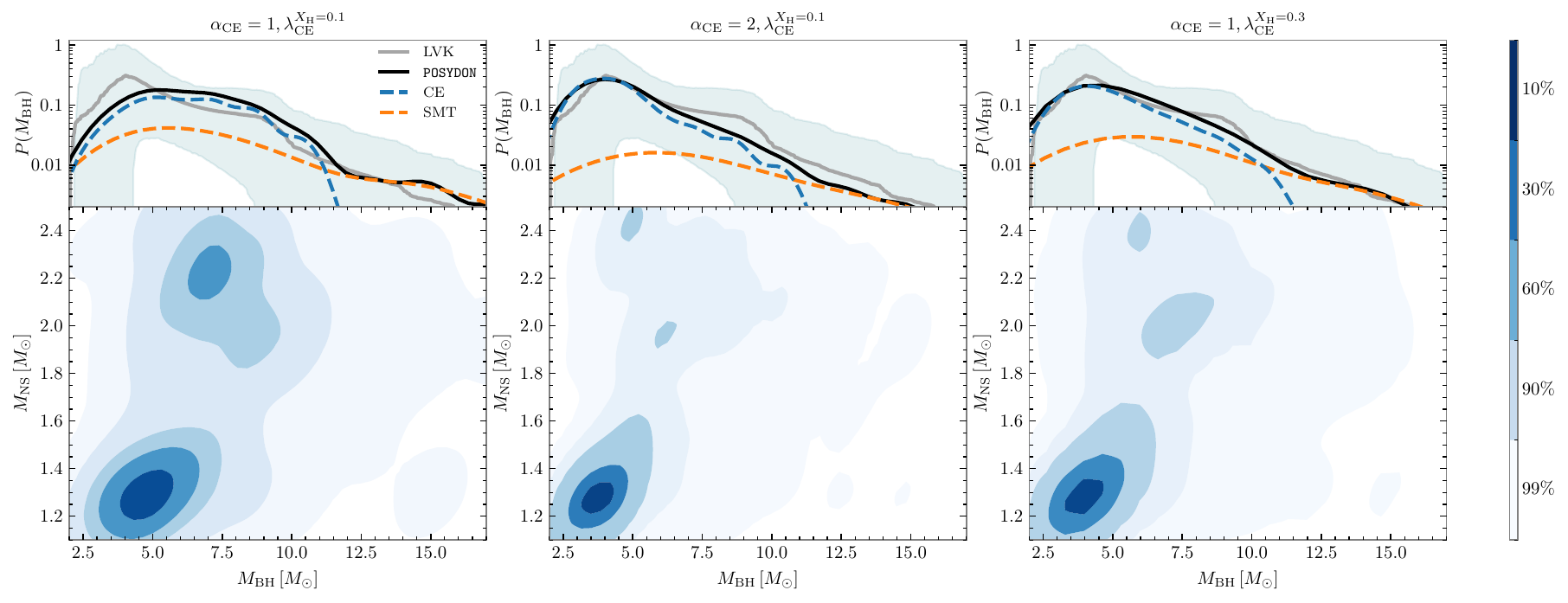}
\caption{NS and BH mass distributions of our intrinsic coalescing NSBH populations. Left and middle panels correspond to the populations generated with CE parameter $\alpha_{\mathrm{CE}} =1$ and $2$, respectively, and the core-envelope boundary is defined at $X_{\rm{H}} = 0.1$. The right panel corresponds to the model with $\alpha_{\mathrm{CE}} =1$ and the core-envelope boundary is defined at $X_{\rm{H}} = 0.3$.} The top panels show the BH mass distributions inferred from the three detected events in gray line with $90\%$ credible interval in teal shaded region and from our population model in black line. The contributions from channel CE and channel SMT are shown using kernel density estimate in blue dashed line and orange dashed line, respectively. The bottom panels show the kernel density estimate plot of component masses ranging from $1\%$ to $90\%$ credible interval region.
\label{fig:alpha}
\end{figure*}

\subsection{The Role of Common Envelope}
In Figure~\ref{fig:alpha}, we show the distribution of BH and NS masses for the intrinsic populations in our simulations, with $\alpha_{\rm{CE}} = 1$ on the left and $\alpha_{\rm{CE}} = 2$ in the middle and the core-envelope boundary is defined at $X_{\rm{H}} = 0.1$. On the right panel, we show the distribution with $\alpha_{\rm{CE}} = 1$ but the core-envelope boundary is defined at $X_{\rm{H}} = 0.3$. Additionally, we present the 1D BH mass distributions and compare them with the distribution inferred from the three events GW230529, GW200105, and GW200115 \citep{2024ApJ...970L..34A}. We also show the contributions from two main formation channels that we define: channel CE, for which the binary experiences CE evolution either before or after the BH formation, and channel SMT, where the binary undergoes at least one stable mass transfer phase without any CE evolution. In channel CE, the classic pathway \citep[e.g.,][]{2006csxs.book..623T} dominates, in which the two non-degenerate stars undergo stable mass transfer, and after the primary star forms a BH, the secondary star and the BH initiate a CE phase. More frequently, the binary experiences a subsequent case~BB mass transfer, in which the secondary helium star expands to fill the Roche lobe due to shell helium burning. In channel SMT, most binaries undergo two stable mass transfer phases, one before and one after the BH formation. Some binaries avoid mass transfer before BH formation and experience CE evolution or stable mass transfer afterward.

Utilizing our default parameter settings, with $\alpha_{\rm{CE}} = 1$ and $X_{\rm{H}} = 0.1$, we can see two prominent peaks: one around $M_{\rm{BH}}\simeq 4-5\,M_{\odot}$ and $M_{\rm{NS}}\simeq 1.3\,M_{\odot}$, and another at $M_{\rm{BH}}\simeq 7-8\,M_{\odot}$ and $M_{\rm{NS}}\simeq 2.1-2.2\,M_{\odot}$. The paucity of NSs around $1.7\,M_{\odot}$ is caused by the discontinuity of $0.1\,M_{\odot}$ in the proto-compact object mass equation at carbon-oxygen core masses of $3.5\,M_{\odot}$ in the \citet{2012ApJ...749...91F} delayed prescription, which was also discussed in \citet{2021MNRAS.508.5028B}. The compact object mass depends not only on the initial progenitor mass, but also on the mass transfer processes during the evolution, which affects the final core mass of the progenitor at core-collapse. Especially for NSs, case~BB mass transfer plays a crucial role, as the amount of mass remaining after this mass transfer phase directly determines the NS mass. Massive NSs typically originate from helium stars that do not lose much mass through case~BB mass transfer or avoid it. If the NS progenitor loses mass such that it ends up with a final helium core in the mass range of $1.4-2.5\,M_{\sun}$ \citep{2004ApJ...612.1044P}, it will form a NS of $\simeq 1.26\,M_{\sun}$ through an ECSN. If the NS progenitor has a final carbon-oxygen core mass below $2.5\,M_{\sun}$, based on the \citet{2012ApJ...749...91F} delayed prescription, it will eventually become a NS of $\simeq 1.27\,M_{\sun}$. We can see from Figure~\ref{fig:alpha} that these light NSs have a significant contribution to mass-gap NSBHs. 

For the default model, the BH mass distribution shows a subdued maximum at $\simeq 5\,M_{\odot}$, stays nearly flat through $\simeq 6-8\,M_{\odot}$, and then drop off into a faint tail beyond $\simeq 10\,M_{\odot}$. Although the simulation curve lies within the $90\%$ credible interval of the BH mass distribution inferred from observations, it does not closely match the mean of the BH mass spectrum. When adopting a larger $\alpha_{\rm{CE}} = 2$, the BH mass distribution shifts towards lower masses and concentrates around $\simeq 4\,M_{\odot}$, predominantly driven by channel CE. This result is consistent with the findings of \citet{2024ApJ...974..211Z}, who found a low-mass peak of $3.4\,M_{\odot}$ within the mass gap when adopting $\alpha_{\rm{CE}} = 2$. We find that a larger $\alpha_{\rm{CE}}$ does not increase the merger rate of NSBHs with heavy BHs ($\gtrsim 6\,M_{\odot}$) but significantly facilitates the formation of NSBHs with light BHs. The reason is that, in the underlying populations, plenty of binaries with light BHs initiate a CE phase and subsequently merge in CE evolution. Compared to binaries containing heavy BHs, they are greater in number but tend to have lower orbital energy, making it harder to eject the envelope of the companion star. Hence, a more efficient energy conversion in CE evolution would substantially increase the number of systems succeeding in CE evolution with light BHs in particular. Additionally, some binaries with heavy BHs would result in wider orbits due to a larger $\alpha_{\rm{CE}}$, and no longer merge within a Hubble time. The fraction of channel CE increases from $\approx 72\%$ for $\alpha_{\rm{CE}} = 1$ to $\approx 87\%$ for $\alpha_{\rm{CE}} = 2$.

Although a larger $\alpha_{\rm{CE}}$ provides a better match to observations by producing a prominent peak within the mass gap, we see a mild deficiency in BHs with masses $\gtrsim 7\,M_{\odot}$. However, in \texttt{POSYDON}, when calculating the parameter $\lambda_{\rm{CE}}$, we include both the gravitational energy and internal energy of the envelope but do not account the recombination energy. As a result, we are inclined to adopt $\alpha_{\rm{CE}}$ values below $1$, as no other energy source is available to justify a large $\alpha_{\rm{CE}}$ of $2$. 

\citet{2019ApJ...883L..45F} and \citet{2022ApJ...937L..42H}, using different approaches, argued that the final post-CE orbital separations maybe larger than predicted by the standard $\alpha - \lambda$ prescription, reminiscing to effective  $\alpha_{\rm{CE}}>1$. In both studies, the large post-CE orbital separations actually originate from the earlier detachment from the rapid insipiral, phase before the donor star is fully stripped, with the remaining envelope being stripped via stable, non-conservative mass transfer. Shifting the core-envelope boundary outward produces similar effect to adopting a larger $\alpha_{\rm{CE}}$,  because it effectively reduces the total binding energy to overcome. Our treatment of CE with the core-envelope boundary defined at $X_{\rm{H}} = 0.3$ is consistent with \citet{2019ApJ...883L..45F}. We can see from Figure~\ref{fig:alpha}, the model with $\alpha_{\rm{CE}} = 1$ and $X_{\rm{H}} = 0.3$ produces a single BH mass peak around $4-5\,M_{\odot}$, similar to the case of $\alpha_{\rm{CE}} = 2$. Moreover, the BH distribution matches the observation well in all mass range.

\subsection{Supernova Natal Kicks}

\begin{figure*}
\includegraphics[width=\textwidth]{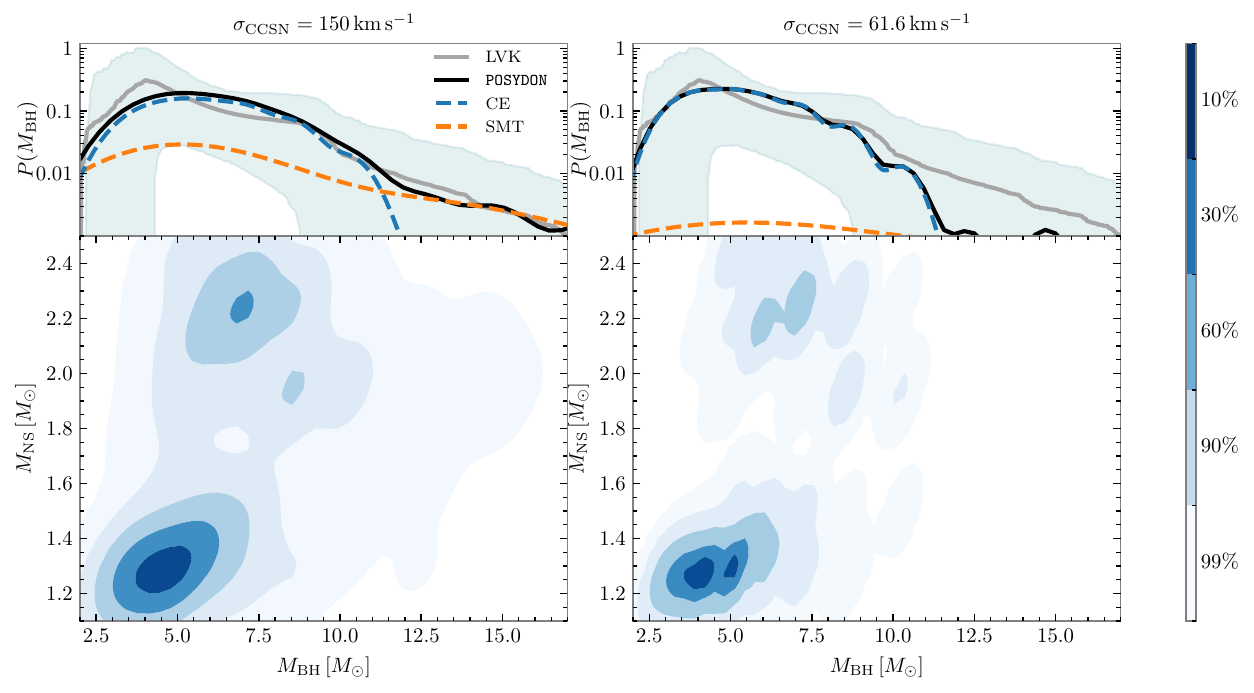}
\caption{Same as Figure~\ref{fig:alpha} with $\alpha_{\rm{CE}}$=1, but generated with different kick velocities of $\sigma_{\mathrm{CCSN}}=150$ and $61.6\,\mathrm{km\,s^{-1}}$.
\label{fig:kicks}}
\end{figure*}

In this section, we investigate the impact of SN kicks on the BH mass distributions of coalescing NSBH binaries. In Figure~\ref{fig:kicks} we show the distribution of BH and NS masses in coalescing NSBH populations for another two different SN kick velocities of $\sigma_{\rm{CCSN}}=150\,\rm{km\,s^{-1}}$ and $\sigma_{\rm{CCSN}}=61.6\,\rm{km\,s^{-1}}$. The populations are generated using $\alpha_{\rm{CE}}=1$ and core-envelope boundary defined at $X_{\rm{H}} = 0.1$. We can see that a lower kick velocity of $\sigma_{\rm{CCSN}}=150\,\rm{km\,s^{-1}}$ does not significantly alter the overall BH mass distribution compared to the default model of $\sigma_{\rm{CCSN}}=265\,\rm{km\,s^{-1}}$ shown in Figure~\ref{fig:alpha}. In the case of $\sigma_{\rm{CCSN}}=61.6\,\rm{km\,s^{-1}}$, the peak shifts toward lower BH masses and channel CE becomes completely dominant. This occurs because the underlying population favors light BHs going through CE evolution, and weak kicks prevent these binaries from being disrupted. Moreover, binaries going through channel SMT typically require high eccentricities induced by strong kicks to merge within a Hubble time. As a result, weak kicks reinforce the dominance of channel CE. The fraction of channel CE is $80\%$ and $98\%$ for $\sigma_{\rm{CCSN}}=150\,\rm{km\,s^{-1}}$ and $\sigma_{\rm{CCSN}}=61.6\,\rm{km\,s^{-1}}$, respectively. 

Although lower kick velocities slightly raise the percentage of lower mass-gap BHs, the overall distribution is still similar to the default model. For $\sigma_{\rm{CCSN}}=61.6\,\rm{km\,s^{-1}}$, the fraction of NSBH mergers with BH masses above $\simeq 11\,M_{\odot}$ becomes marginal. Furthermore, this low kick velocity increases the local merger rate density of NSBHs to an excessively high value. We estimate the local merger rate density of NSBH mergers for different kick velocities at redshift zero. The three kick velocities $\sigma_{\rm{CCSN}}=265\,\rm{km\,s^{-1}}$, $\sigma_{\rm{CCSN}}=150\,\rm{km\,s^{-1}}$, and $\sigma_{\rm{CCSN}}=61.6\,\rm{km\,s^{-1}}$ correspond to local merger rate densities of $86\,\rm{Gpc^{-3}\,yr^{-1}}$, $247\,\rm{Gpc^{-3}\,yr^{-1}}$, and $478\,\rm{Gpc^{-3}\,yr^{-1}}$, respectively. The rate for the lowest kick velocity, $\sigma_{\rm{CCSN}}=61.6\,\rm{km\,s^{-1}}$, is significantly higher than the upper limit of the NSBH local merger rate estimated from LVK analysis.

\subsection{Account for the Mass-Gap Black Holes in Low-Mass X-ray Binaries}

\begin{figure*}
\includegraphics[width=\textwidth]{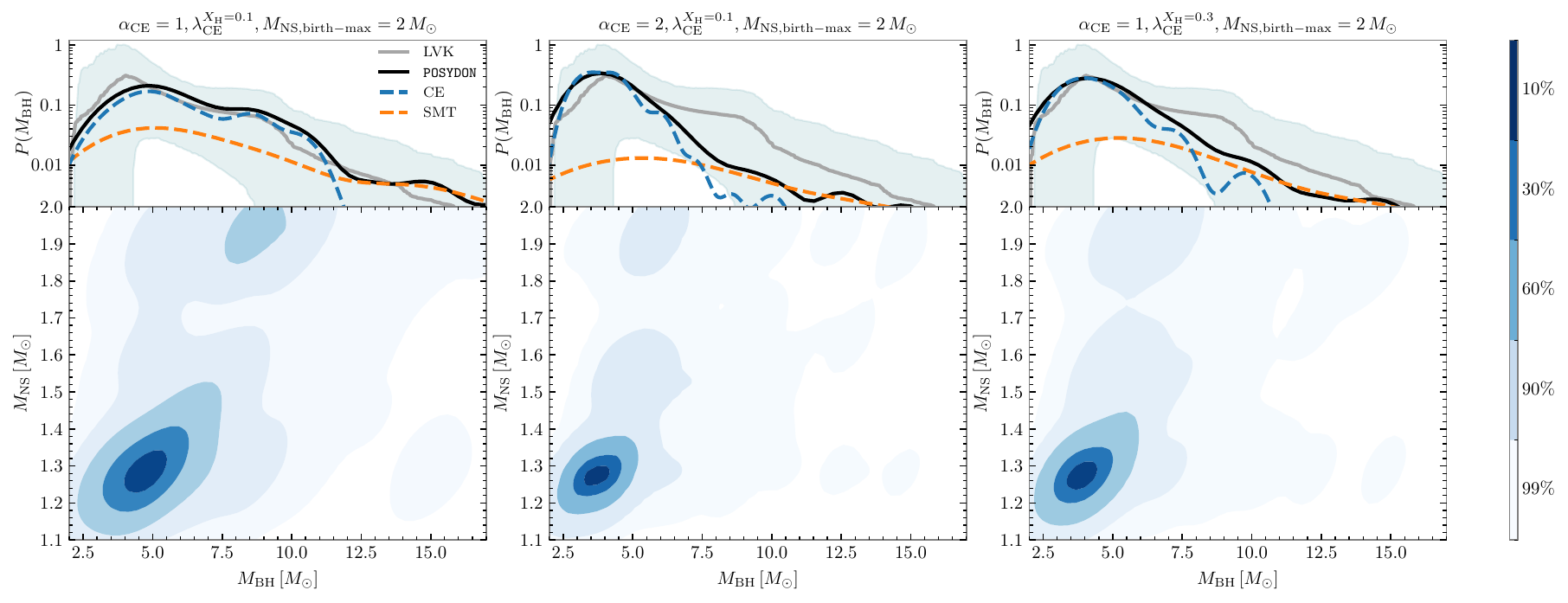}
\caption{Similar to Figure~\ref{fig:alpha} but assuming the maximum NS birth mass is $2\,M_{\odot}$. 
\label{fig:nsbirth_ce}}
\end{figure*}

\begin{figure*}
\includegraphics[width=\textwidth]{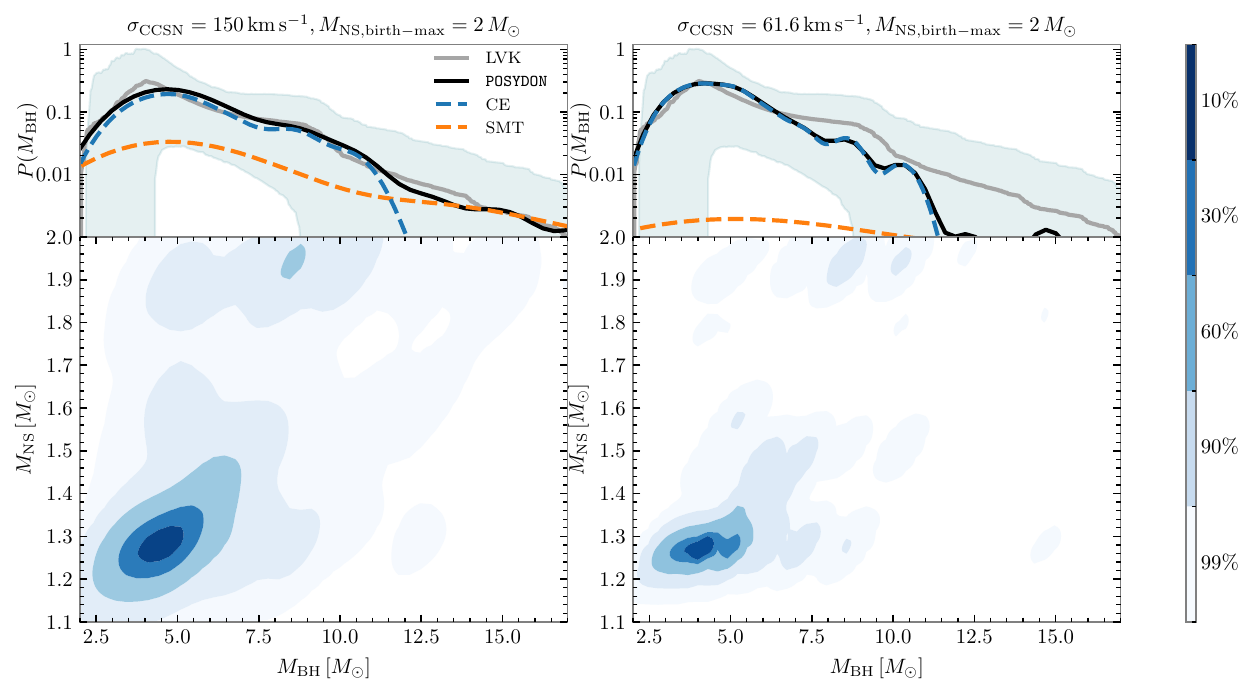}
\caption{Similar to Figure~\ref{fig:kicks} but assuming the maximum NS birth mass is $2\,M_{\odot}$. 
\label{fig:nsbirth_kick}}
\end{figure*}

In this section, we consider the constraints on the maximum NS birth mass. Following the investigation of selection biases against mass-gap BHs in the LMXB population \citep{2023ApJ...954..212S}, we exclude binaries with NS birth masses exceeding the maximum value of $2\,M_{\odot}$ from our intrinsic population. Figure~\ref{fig:nsbirth_ce} and Figure~\ref{fig:nsbirth_kick} shows the distributions of NS and BH masses from the same populations presented in Figure~\ref{fig:alpha} and Figure~\ref{fig:kicks}, but with the maximum NS birth mass set to $2\,M_{\odot}$. With this restriction, we can see that the default setting with $\alpha_{\rm{CE}}=1$ and $\sigma_{\rm{CCSN}}=265\,\rm{km\,s^{-1}}$ also predicts a prominence of BH masses around the high-end edge of the lower mass gap. The number of high-mass BHs around $7-9\,M_{\odot}$ is reduced, as most of them are paired with high-mass NSs above $2\,M_{\odot}$, which are removed from the population. Now, a higher $\alpha_{\rm{CE}}=2$ predicts an overabundance of mass-gap BHs, particularly those with masses $\lesssim 4\,M_{\odot}$, and a paucity of BHs with mass above $\simeq 7\,M_{\odot}$. Less than $\approx 10\%$ of coalescing NSBH systems contain BHs outside the mass gap. Same trend occurs for the model with the core- envelope boundary define at $X_{\rm{H}} = 0.3$, but it is less significant. Both models with lower kick velocities predict a BH mass peak within $4-5\,M_{\odot}$. For the moderate kick velocity $\sigma_{\rm{CCSN}}=150\,\rm{km\,s^{-1}}$, the BH mass distribution also closely aligns with the mean of the BH mass spectrum inferred from observations.

The local merger rate densities of NSBHs and mass-gap NSBHs for all the model variations are summarized in Table~\ref{rate}. With the restriction on NS birth mass, $\mathcal{R}_{\rm{gap}}$ barely changes, as mass-gap BHs are accompanied by low-mass NSs. With $M_{\rm{NS,birth-max}} = 2\,M_{\odot}$, the three kick velocities $\sigma_{\rm{CCSN}}=265\,\rm{km\,s^{-1}}$, $\sigma_{\rm{CCSN}}=150\,\rm{km\,s^{-1}}$, and $\sigma_{\rm{CCSN}}=61.6\,\rm{km\,s^{-1}}$ correspond to local merger rate densities of $40\,\rm{Gpc^{-3}\,yr^{-1}}$, $119\,\rm{Gpc^{-3}\,yr^{-1}}$, and $274\,\rm{Gpc^{-3}\,yr^{-1}}$, respectively. The lowest kick velocity $\sigma_{\rm{CCSN}}=61.6\,\rm{km\,s^{-1}}$ still predicts a merger rate density exceeding the upper limit of the updated rate inferred from the three events. This is reasonable as the velocity should be considered as a lower limit \citep{2023MNRAS.521.2504O}. In the other two cases, the predicted merger rates are consistent with the inferred merger rates.

We note that in \citet{2024ApJ...974..211Z}, the merger rate densities predicted are much lower than ours. Moreover, their merger rate densities barely change when $\alpha_{\mathrm{CE}}$ is increased from $1$ to $2$. The primary reason for these differences is that they adopted the ``pessimistic'' approach \citep{2012ApJ...759...52D} for CE evolution, which assumes that CE evolution with a donor star on the Hertzsprung gap, as classified by the SSE formulae \citep{2000MNRAS.315..543H} always leads to a merger, postulating that the lack of a steep density gradient near the core-envelope boundary of these stars will prevent them from detaching successfully from a CE phase. This ad-hoc criterion for the outcome of CE phases was adopted to limit the high efficiency of the CE channels predicted by many rapid BPS codes. However, detailed binary evolution models suggest that the latter is likely related to the very conservative mass-transfer stability criteria implemented in those codes \citep[see, e.g.,][]{2021ApJ...922..110G,2021A&A...650A.107M}.

\subsection{Observable Population}
\begin{figure*}
\hspace*{3cm}\includegraphics[scale=0.5]{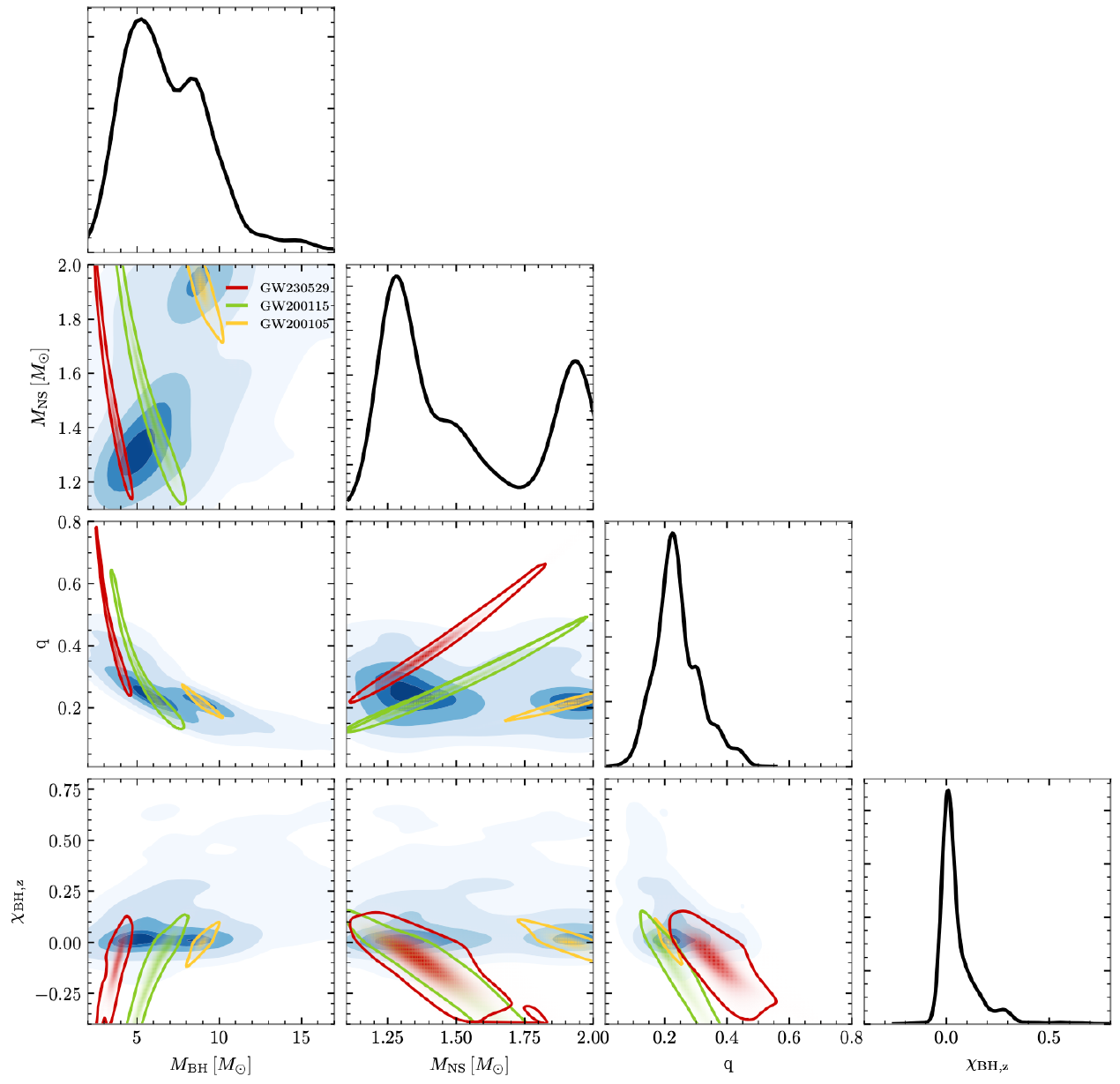}
\caption{Corner plot of the observable population of coalescing NSHBs in our simulation, generated with $\alpha_{\mathrm{CE}} =1$, $\sigma_{\mathrm{CCSN}}=265\,\rm{km\,s^{-1}}$ and $M_{\rm{NS,birth-max}} = 2\,M_{\odot}$. We show the BH mass, NS mass, mass ratio, and BH spin component parallel to the direction of the orbital plane. The contour colors represent the same percentage intervals as in the previous figures. The red, green, and yellow contours indicate the $90\%$ confidence interval of GW230529 \citep{2024ApJ...970L..34A}, GW200115, and GW200105 \citep{2021ApJ...915L...5A}, respectively.
\label{fig:obs}}
\end{figure*}

In the previous sections, we presented BH and NS mass distributions of the intrinsic coalescing NSBH populations in our simulations, which are not directly comparable to observed events. In Figure~\ref{fig:obs}, we display the properties of the observable population for the model with $\alpha_{\mathrm{CE}} =1$, $X_{\rm{H}} = 0.1$ for the core-envelope boundary, $\sigma_{\rm{CCSN}}=150\,\rm{km\,s^{-1}}$, and $M_{\rm{NS,birth-max}} = 2\,M_{\odot}$, which produces a BH mass distribution close to the BH mass spectrum inferred from observations. The corner plot includes the distributions of BH mass, NS mass, mass ratio, and the BH spin at the direction of the orbit $\chi_{\rm{BH,z}}$, as these same quantities for GW230529 were shown in \citet{2024ApJ...970L..34A}. We also display the posterior distributions of the three observed events GW230529 \citep{2024ApJ...970L..34A}, GW200105, and GW200115 \citep{2021ApJ...915L...5A} as contours representing the $90\%$ credible regions. Our simulation matches the properties of the three events well. A higher detection probability for BHs around $\simeq 8\,M_{\odot}$ and NSs around $\simeq 1.9\,M_{\odot}$ appears, which is consistent with the event GW200105, with respect to the intrinsic population. The mass ratio distribution centers at $\simeq 0.2$ and rarely exceeds $0.5$. The $\chi_{\rm{BH,z}}$ is centered at zero, with a spread to $\simeq 0.2$ and a small bump at $\simeq 0.25$. Most BHs have negligible natal spins because BH progenitors lose most of their angular momentum through mass transfer and winds during the evolution. The small bump corresponds to initially close binaries where tides can spin up the BH progenitors, and these initially close binaries typically go through channel SMT \citep{2024A&A...683A.144X}. Because we assume that BH accretion is Eddington-limited, the accretion spin-up is not significant. The BHs can be spun up by case~BB mass transfer but the increase is typically below $0.1$. BH spins as high as $\simeq 0.5$ are attributed to stable case~A mass transfer, but the fraction of such cases is negligible. The three events are consistent with nearly zero $\chi_{\rm{BH,z}}$. For GW230529, if its primary compact object is a massive NS, as the mass ratio is anti-correlated with the primary mass, the secondary NS is expected to be massive at birth, even exceeding the threshold of $2\,M_{\odot}$. If its primary is a BH, it is more likely that the BH mass is at the higher end.

\begin{deluxetable*}{lcc}
\tablecaption{Local Merger Rate Density of NSBH Mergers for Different Models \label{rate}}
\tablehead{\colhead{Model}& \colhead{$\mathcal{R}_{\rm{NSBH}} [\rm{Gpc}^{-3}\rm{yr}^{-1}]$} & \colhead{$\mathcal{R}_{\rm{gap}} [\rm{Gpc}^{-3}\rm{yr}^{-1}]$} }

\startdata
default& $86$ & $12$ \\
$\alpha_{\rm{CE}} =2$ & $184$ & $64$ \\
$\lambda_{\rm{CE}}^{X_\mathrm{H}=0.3}$ & $110$ & $30$ \\
$\sigma_{\rm{CCSN}}=150\,\rm{km\,s^{-1}}$ & $247$ &$46$ \\
$\sigma_{\rm{CCSN}}=61.6\,\rm{km\,s^{-1}}$ & $478$ &$124$ \\
$M_{\rm{NS,birth-max}} = 2\,M_{\odot}$ & $55$ &$12$ \\
$\alpha_{\rm{CE}} =2$, $M_{\rm{NS,birth-max}} = 2\,M_{\odot}$ & $130$ &$61$ \\
$\lambda_{\rm{CE}}^{X_\mathrm{H}=0.3}$, $M_{\rm{NS,birth-max}} = 2\,M_{\odot}$ & $73$ & $28$ \\
$\sigma_{\rm{CCSN}}=150\,\rm{km\,s^{-1}}$, $M_{\rm{NS,birth-max}} = 2\,M_{\odot}$ & $155$ &$45$ \\
$\sigma_{\rm{CCSN}}=61.6\,\rm{km\,s^{-1}}$, $M_{\rm{NS,birth-max}} = 2\,M_{\odot}$ & $340$ &$120$ \\
\enddata
\tablecomments{Inferred rates from LVK: $\mathcal{R}_{\rm{NSBH}}= 94^{+109}_{-64}\,\rm{Gpc}^{-3}\rm{yr}^{-1}$ and $\mathcal{R}_{\rm{gap}}= 24^{+28}_{-16}\,\rm{Gpc}^{-3}\rm{yr}^{-1}$ or $33^{+89}_{-29}\,\rm{Gpc}^{-3}\rm{yr}^{-1}$ \citep{2024ApJ...970L..34A}.\\
default: $\alpha_{\rm{CE}} =1$, $\lambda_{\rm{CE}}^{X_\mathrm{H}=0.1}$,$M_{\rm{NS,birth-max}} = 2.5\,M_{\odot}$, $\sigma_{\rm{CCSN}}=265\,\rm{km\,s^{-1}}$.}

\end{deluxetable*}


\subsection{Electromagnetic Counterparts}
To estimate the fraction of potential associated EMCs in the coalescing NSBH population, we utilize the empirical formula provided by \citet{2018PhRvD..98h1501F} to calculate the remnant mass outside the innermost stable circular orbit (ISCO) of the BH resulting from the tidal disruption of the NS. Systems having nonzero remnant mass outside the ISCO are considered to produce EMCs, such as short gamma-ray bursts and kilonovae. We assume three NS radii $R_{\mathrm{NS}} = 11\,\rm{km}$, $12\,\rm{km}$, and $13\,\rm{km}$, simply representing a range of NS EoS from soft to stiff. For the default model with $\alpha_{\mathrm{CE}} =1$, $X_{\rm{H}} = 0.1$ for the core-envelope boundary, and $\sigma_{\mathrm{CCSN}}=265\,\rm{km\,s^{-1}}$, the EMC fractions for the intrinsic population are $5\%$, $22\%$, and $34\%$ for $R_{\mathrm{NS}} = 11\,\rm{km}$, $12\,\rm{km}$, and $13\,\rm{km}$, respectively. Under the variation of $X_{\rm{H}} = 0.3$ for the core-enevelope boundary definition, the EMC fractions increase to $23\%$, $56\%$, and $59\%$. In the model with a maximum NS birth mass constraint of $2\,M_{\odot}$, the fractions are $7\%$, $33\%$, and $50\%$ for $\sigma_{\mathrm{CCSN}}=265\,\rm{km\,s^{-1}}$ and $8\%$, $38\%$, and $59\%$ for $\sigma_{\mathrm{CCSN}}=150\,\rm{km\,s^{-1}}$. The EMC fractions increase substantially with an outer core-envelope boundary and with the NS birth mass limit because the models produces a higher proportion of systems containing light BHs and light NSs, which raises the probability of tidal disruption for the NSs outside the ISCO.   

Shifting from the intrinsic to the observable populations, we estimate the EMC fractions for detectable systems. The default model yields EMC fractions of $1\%$, $7\%$, and $12\%$ for $R_{\mathrm{NS}} = 11\,\rm{km}$, $12\,\rm{km}$, and $13\,\rm{km}$, respectively. The model of $X_{\rm{H}} = 0.3$ for the core-envelope boundary increase the fractions to $7\%$, $20\%$, and $23\%$. With the limit on the NS birth mass, we find EMC fractions of $2\%$, $14\%$, and $25\%$ for $\sigma_{\mathrm{CCSN}}=265\,\rm{km\,s^{-1}}$ and $3\%$, $20\%$, and $32\%$ for $\sigma_{\mathrm{CCSN}}=150\,\rm{km\,s^{-1}}$. The detection of GW230529 shifts the BH mass distributions toward lower masses in the coalescing NSBH population, leading to an increased chance of observing associated EMCs, especially if the NS EoS is stiff. 

\section{Discussion} \label{sec:discussion}

\subsection{Uncertainties}

One of the primary uncertainties in our BPS study is the SN prescription, specifically the association between pre-SN progenitor core masses and the compact object masses. The \citet{2012ApJ...749...91F} delayed prescription allows for forming compact objects within the lower-mass gap. However, the discontinuity of NS mass around $1.7\,M_{\odot}$ is neither supported by the observations nor motivated by SN physics. Furthermore, the restriction on the maximum NS birth mass cannot be reconciled self-consistently with the \citet{2012ApJ...749...91F} delayed prescription, suggesting a potential need for a different relation between heavy pre-SN core masses and compact object masses. Hydrodynamics simulations have been used to study the fallback and the remnant properties for core-collapse SN \citep{2018ApJ...852L..19C, 2020MNRAS.495.3751C, 2021ApJ...920L..17V}. \citet{2021ApJ...920L..17V} proposed that a massive stripped helium star can lead to a massive NS or a low-mass BH depending on the explosion energy. \citet{2020MNRAS.495.3751C} showed that an $40\,M_{\odot}$ zero-metallicity star can form a mass-gap BH with enhanced explosion energy in their simulation. However, a comprehensive simulation-based prescription that links the progenitor's helium or carbon-oxygen core properties to the outcome compact object remnant masses has not been established. Such a prescription is crucial for accurately modeling BH and NS mass distributions in population studies of coalescing NSHBs and for determining the fraction of associated EMCs. In addition, the conditions that trigger ECSNe remain highly uncertain. \citet{2015MNRAS.451.2123T} identified a very narrow range of pre-SN core masses for ECSNe. The initial mass range for ECSNe is affected by uncertainties in mass loss during the late stages of stellar evolution and different treatments of stellar mixing and convection \citep{2013ApJ...771...28T,2015ApJ...810...34W,2017ApJ...850..197P}. Compared to \citet{2004ApJ...612.1044P}, these studies proposed a narrower initial mass range for ECSNe in binaries. Future observations of double compact object mergers will put further constraints on various SN prescriptions and advance our understanding of SN physics.

Models of metallicity- and redshift-dependent SFH can affect primarily the rates, and perhaps modestly alter the shapes of NSBH property distributions \citep{2021MNRAS.508.5028B}. We leave the exploration of these factors to future studies when the data sample grows large enough for aiding to allow a proper statistical comparison. Furthermore, investigating variations in stellar physics like the wind prescriptions and overshooting parameters, and in binary physics like the mass transfer efficiency is beyond the scope of this study, as changing these model assumptions would require constructing new binary simulation grids. However, the \texttt{POSYDON} framework allows for potential future investigations into these key physics self-consistently and precisely by integrating new grids of detailed binary models.

\subsection{Maximum Neutron Star Birth Mass}

Several massive NSs have been discovered with mass exceeding $2\,M_{\odot}$, for example 
PSR J0740+6620 with $M_{\rm{NS}}=2.08\pm{0.07}\,M_{\odot}$ \citep{2021ApJ...915L..12F}, J1810+1744 with $M_{\rm{NS}}=2.13\pm{0.04}\,M_{\odot}$ \citep{2021ApJ...908L..46R}, and PSR J0952-0607 with $M_{\rm{NS}}=2.35\pm{0.17}\,M_{\odot}$ \citep{2022ApJ...934L..17R}. All of these NSs are millisecond pulsars with a low-mass main-sequence or a white dwarf companion. They are expected to have accreted mass from their companion and been recycled. As a result, their birth masses remain unknown and could be less than $2\,M_{\odot}$. In contrast, the majority of non-recycled NSs in the Galaxy have low masses around $1.3\,M_{\odot}$ \citep{2012ApJ...757...55O,2013ApJ...778...66K}. In addition, the NS candidates in wide binaries from Gaia DR3 have masses around $1.4\,M_{\odot}$, with none exceeding $2\,M_{\odot}$ \citep{2024OJAp....7E..58E}. In the population of eclipsing NS high-mass X-ray binaries, the NSs are very young and are not expected to have accreted a significant amount of mass. In the study by \citet{2015A&A...577A.130F}, most NSs in high-mass X-ray binaries are measured to be below $2\,M_{\odot}$, with only one possible exception of Vela X-1, which contains a NS with mass $M_{\rm{NS}}=2.12\pm{0.16}\,M_{\odot}$. A recent study examined the birth mass function of NSs by applying probabilistic corrections to mass measurements of $90$ NSs. They found a peak at $\simeq 1.27\,M_{\odot}$ and a steep decline following a power-law trend toward higher masses, suggesting that NSs with birth masses exceeding $2\,M_{\odot}$ are likely rare \citep{2024arXiv241205524Y}.

\subsection{Could the NS Form First?}

\citet{2024A&A...683A.144X} found that the BH always forms first in the population of coalescing NSBHs at solar metallicity from \texttt{POSYDON} simulations. This is because the rotation-dependent accretion model for non-degenerate stars in \texttt{POSYDON} suggests a low accretion efficiency for case~B mass transfer in general, which limits the probability of mass reversal. Other population studies have similarly found that low accretion efficiency for non-degenerate accretors leads to a low fraction of first-born NSs in NSBH populations \citep{2018MNRAS.481.1908K,2021ApJ...920...81S,2021MNRAS.508.5028B}. In contrast, nearly completely conservative mass transfer can yield a higher fraction of first-born NSs in NSBHs \citep{2021ApJ...912L..23R} and facilitate the formation of millisecond pulsar BH binaries, where the NSs form prior to the BHs to be recycled by the BH progenitors \citep{2021MNRAS.504.3682C}. \citet{2024arXiv241215521L} also finds that \texttt{POSYDON} predicts a low formation rate of first-born NSs in NSBHs and that millisecond pulsar BH binaries cannot form. In the \texttt{POSYDON} grids, rejuvenation may occur in some close binaries, where tidal forces spin down the accretor and enhance mass accretion. However, the first-born NSs are highly likely to enter a CE phase with massive BH progenitors and subsequently merge due to their low orbital energy \citep[see a detailed discussion in][]{2024A&A...683A.144X}. In this study, we find that this conclusion also holds true for varied metallicities. 

We posit that if the accretion efficiency is actually high enough to lead to mass reversal for binaries with asymmetric masses in wide orbits, the fraction of first-born NSs in coalescing NSBHs can be nontrivial. \citet{2021ApJ...912L..23R} predicted that the fraction of NSBH mergers in which the NS forms first can reach about $10\%$ when assuming that accretion onto non-degenerate stars is limited by the thermal timescale of the accetor, which yields an accretion efficiency much higher than that predicted by the rotation-dependent accretion model. Similarly, \citet{2021MNRAS.504.3682C} made the same assumption regarding mass accretion and found that NSs can form first and be recycled during the subsequent CE evolution and case~BB mass transfer phase. However, the NSs are not expected to gain a significant amount of mass during CE evolution and case~BB mass transfer from the BH progenitors. \citet{2024arXiv241215521L} found that a small group of NSBHs with first-born NSs form through double CE evolution between two non-degenerate stars. However, no mass transfer occurs subsequently because the BH progenitors are too massive to expand sufficiently to fill their Roche lobes. In conclusion, we expect that even if the NS forms first in coalescing NSBHs, its mass is likely to be close to its birth mass.


\section{Conclusion} \label{sec:conclusion}

The detection of the new event GW230529, which very likely contains a compact object with mass in the lower mass gap, has refreshed our understanding of the coalescing NSBH populations. The inclusion of GW230529 in population inference analyses yields a new BH mass distribution for coalescing NSBHs and an updated merger rate. Most importantly, it suggests that the lower mass gap may not exist. We conduct a population synthesis study on coalescing NSBHs to investigate how the CE process and SN kick velocity affect their properties, and to reconcile the detection of mass-gap BHs from GWs with their absence in LMXBs.

We find that, using the \texttt{POSYDON} default model with $\alpha_{\rm{CE}} = 1$, $X_{\rm{H}} = 0.1$ for the core-envelope boundary, $\sigma_{\mathrm{CCSN}}=265\,\rm{km\,s^{-1}}$, and $M_{\rm{NS,birth-max}} = 2.5\,M_{\odot}$, the BH mass distribution does not closely match the BH mass spectrum inferred from the three events GW230529, GW200105, and GW200115, showing a lower fraction of mass-gap BHs. Increasing the CE efficiency to $\alpha_{\rm{CE}} = 2$ can efficiently facilitate the formation of NSBH mergers with mass-gap BHs. These light BHs going through CE are greater in number in the underlying population, but they are hard to survive CE due to low orbital energy. A higher CE efficiency would increase the success rate of CE for these light BHs, and thus lead to a surge of coalescing NSBH with mass-gap BHs. However, a large CE efficiency of $\alpha_{\rm{CE}} = 2$ is hard to justify as most energy sources have been considered in our calculation. Alternatively, we find that an outer boundary of the core and the envelope produces similar effects. Adopting $X_{\rm{H}} = 0.3$ to define the core-envelope boundary, which is motivated by \citet{2019ApJ...883L..45F}, we find that the simulated BH mass distribution matches the observations well. Furthermore, we find that lower kick velocities can increase the contribution of mass-gap BHs that go through CE in coalescing NSBHs. However, varying kick velocities has limited effects on the overall BH mass distribution.

To explain the selection bias against mass-gap BHs in LMXBs, \citet{2023ApJ...954..212S} found that a maximum NS birth mass of $\simeq 2\,M_{\odot}$ should be considered. With the restriction of NS birth mass, we find that a large $\alpha_{\rm{CE}}$ is not necessary for explaining the updated BH mass distribution from the analysis including GW230529. In this case, a kick velocity of $\sigma_{\mathrm{CCSN}}=265\,\rm{km\,s^{-1}}$ and $\sigma_{\mathrm{CCSN}}=150\,\rm{km\,s^{-1}}$ can better match the BH mass distribution and predict reasonable local merger rates for coalescing NSBHs and mass-gap NSBHs. A lower kick of $\sigma_{\mathrm{CCSN}}=61.6\,\rm{km\,s^{-1}}$, obtained from NS binary systems as a lower limit, leads to an overly high merger rate. 

Note that all our simulation models produce BH mass spectrum within the $90\%$ credible interval of the inferred BH mass distribution, which has large uncertainties due to the small sample size of only three events. In this work, the major comparable feature of the BH mass distribution refers to the peak within the mass gap, which appears after the inclusion of GW230529 in the analysis. Therefore, the conclusions that can be drawn regarding model preferences based on the shape of the BH mass distribution has limitations at this stage. However, the intrinsic BH mass distribution of coalescing NSBHs will becoming increasingly well constrained as the sample size increases. Future detections of NSBH merger events will gradually reveal the physical features explored in this study with increasing confidence.

We also show that the properties of the three NSBH merger events are consistent with our observable populations. The isolated binary evolution channel predicts low BH spins in general. As a result, assuming it formed through the isolated binary evolution channel, the primary star mass of GW230529 can be well constrained to be located within the lower mass gap at the higher end. Moreover, the detection of GW230529 results in an increase in the EMC fraction of coalescing NSBHs predicted by population models especially if the NS equation of state is stiff, as light BHs and NSs are easier to result in tidal disruption. 

Our study has illustrated that coalescing NSBH population is valuable for understanding SN mechanisms, CE evolution, and NS physics. Future detections of NSBH merger events could reveal whether the NS birth mass is limited, provide insights into SN remnant mass prescriptions, and further constrain CE physics and NS EoS.

\section*{acknowledgments}
This work was supported by the Swiss National Science Foundation (project number PP00P2\_211006 ).
The POSYDON project is supported primarily by two sources: the Swiss National Science Foundation (PI Fragos, project numbers PP00P2\_211006 and CRSII5\_213497) and the Gordon and Betty Moore Foundation (PI Kalogera, grant awards GBMF8477 and GBMF12341). 
V.K.\ was partially supported through the D.I.Linzer Distinguished University Professorship fund.
Z.X. acknowledges support from the China Scholarship Council (CSC) and CIERA.
E.Z. acknowledges funding support from the Hellenic Foundation for Research and Innovation (H.F.R.I.) under the "3rd Call for H.F.R.I. Research Projects to support Post-Doctoral Researchers" (Project No: 7933). 
M.B. acknowledges support from the Boninchi Foundation, the project number CRSII5\_21349, and the Swiss Government Excellence Scholarship.
K.A.R.\ is also supported by the Riedel Family Fellowship and thanks the LSSTC Data Science Fellowship Program, which is funded by LSSTC, NSF Cybertraining Grant No.\ 1829740, the Brinson Foundation, and the Moore Foundation; their participation in the program has benefited this work.
K.K. acknowledges support from the Spanish State Research Agency, through the María de Maeztu Program for Centers and Units of Excellence in R\&D, No. CEX2020-001058-M. 
J.J.A.~acknowledges support for Program number (JWST-AR-04369.001-A) provided through a grant from the STScI under NASA contract NAS5-03127. 

%

\vspace{5mm}






\bibliography{sample631}{}
\bibliographystyle{aasjournal}



\end{document}